\documentclass[twocolumn]{emulateapj}
\newcommand{\snIa}{\mbox{SN\,{\sc{}i}a}}

\begin{document}

\title{Probabilistic Cross-Identification of Cosmic Events}
\journalinfo{Accepted for publication in the ApJ}

\author{Tam\'as Budav\'ari}
\affil{Dept.\ of Physics and Astronomy, The Johns Hopkins University, 3400 North Charles Street, Baltimore, MD 21218; budavari@jhu.edu}

\shortauthors{Tam\'as Budav\'ari}
\shorttitle{Cross-Identification of Events}

\begin{abstract} 
We discuss a novel approach to identifying cosmic events in separate and independent observations. In our focus are the true events, such as supernova explosions, that happen once, hence, whose measurements are not repeatable. Their classification and analysis have to make the best use of all the available data. Bayesian hypothesis testing is used to associate streams of events in space and time. Probabilities are assigned to the matches by studying their rates of occurrence. A case study of Type {\sc{}i}a supernovae illustrates how to use lightcurves in the cross-identification process. Constraints from realistic lightcurves happen to be well-approximated by Gaussians in time, which makes the matching process very efficient. Model-dependent associations are computationally more demanding but can further boost our confidence.
\end{abstract}

\email{budavari@jhu.edu}

\section{Introduction} 
\label{sec:intro}
\noindent
Advancements in detector technology have recently opened up new opportunities for time-domain astronomy. The next-generation surveys will observe most of the sky not once but many times over, which enables for systematic studies of variable and transient sources.
The first such programs are already online.
One example is the Catalina Real-Time Transient Survey \citep[CRTS;][]{crts} that uses three dedicated telescopes, and has reported over two thousand new transients to date. Another is the Palomar Transient Factory \citep[PTF;][]{ptf} that has discovered close to 800 supernovae.
The Pan-STARRS telescope 
that started survey operation in 2010, will see thousands of supernovae a month, not to mention the SkyMapper project \citep{skymapper} or the Large Synoptic Survey Telescope (LSST) 
that aims to make the ``greatest movie of the Universe'' by covering most of the sky in every few days.

Modern surveys implement automated discovery in their pipelines. They provide live streams of events in both machine and human readable formats to facilitate rapid follow-up observations and real-time analysis.
An interoperable representation is the VOEvent format defined by the International Virtual Observatory Alliance (IVOA, see \url{www.ivoa.net}).
Capitalizing on this standard, the \url{VOEventNet.org} project promises to deliver alerts to interested subscribers within minutes of discovery.
Another important reason for standardized live streams is the joint analysis of the events. Users of the \url{SkyAlert.org} site can browse for discoveries in a growing selection of data streams. The project will automatically associate any events within 10 degrees and less than 1 hour apart (Roy Williams, private communication).

One outstanding issue is the proper statistical foundation of event aggregation.
%
%
The methods developed for cross-identification of catalogs are not directly applicable. While positional information is clearly needed for transient events, too, one cannot statistically interpret the spatial thresholds. To make a probabilistic statement one has to consider the density but the host objects might not be visible, only the explosions.
In fact, events really are more abstract entities in spacetime, whose density we have to study in 3+1 dimensions.
Rigorous statistical assessments are most important in the presence of crowding: large density compared to the uncertainties.
Spatially the optical observations are typically much more accurate than the X-ray, radio, let alone gamma or gravitational wave detections.
In time, this is especially relevant considering that the frequency of events depends on the cutoff parameter of the detection. By lowering the threshold we can study fainter and more distant events that are although low significance in any one survey, could be interesting when seen in many. In this paper, our goal is to derive the probabilistic matching formalism for transient events, which enables the automated association of detections in space and time.

In Section~\ref{sec:bf} we use Bayesian hypothesis testing to evaluate the quality of candidate associations.
Section~\ref{sec:prob} describes why it is {\em{}not} impossible after all to derive probabilities for these matches.
In Section~\ref{sec:adv} we analyze simulated \snIa{} lightcurves to derive time constraints, and show an accurate approximation that provides analytic results. 
Section~\ref{sec:sum} concludes our study. 
Throughout the paper, the capital $P$ and $L$ symbols represent probabilities and likelihoods, respectively, and the lower-case $p$ letters are probability densities.

\section{Matching Events} 
\label{sec:bf}
\noindent 
Many things happen in the sky. Asteroids move, stars pulsate, supernovae and gamma-ray bursts explode all over the place.
Such astronomical events are typically discovered in photometric images based on the change in the apparent brightness between different epochs.
The imaging cadence is strategically planned to meet the scientific goals. The epochs are optimized for the characteristic time-scale of the intended targets, which might be weeks, days, or just tens of minutes.
Looking at repeated observations, we can spot the significant changes, e.g., using a 5$\sigma$ threshold. At that point, it is possible go back to extract the lightcurve of the detected event from the earlier images.
Variable sources are revisited several times over the course of a survey, which provides ample amount of data. Cosmic events, however, happen just once and we need to make the best use of the observations. 

Considering that different photometric pass-bands potentially probe different physical processes, the lightcurves can be quite a bit different. The detections can be delayed from one instrument to another.
One thing, however, is common: the start time of the event, which can be estimated in all detections by looking at the changing flux as function of time.

Let us now consider a single candidate association that consists of $n$ reported events, one from each survey $i$ (where $i$=1 to $n$). They will have measured \mbox{$D_x\!=\!\{x_i\}$} positions in the sky (unit vectors of the directions) and can constrain \mbox{$D_t\!=\!\{t_i\}$} start times with some uncertainties.
%
%
The time constraint is not the precision of our clock, but rather the ability of our model $M$ of the given survey to estimate the start time $t$ of a particular event. 
Formally, it is a likelihood $L(\tau|t)\!\equiv\!p(t|\tau,M)$ that is a function of the model's $\tau$ epoch.
For now, let us assume that $L_i(\tau|\cdot)$ is known for every survey. 
%
In \mbox{Section\,\ref{sec:adv}} we take a closer look at how to derive such likelihood functions from observed lightcurves; in particular we analyze the Type {\sc{}i}a supernovae in detail.

We ask the question whether the detections in the given tuple are from the same event or not using Bayesian hypothesis testing.
We wish to compare two hypotheses.
The first one, call it $H$, claims that all detections are of the same cosmic event, hence they have a common epoch ($\tau$) and a common true position (both unknown to us.)
The second hypothesis is the former's complement $K$ that allows for any one of the detections to be from a separate event, which we model using a set of parameters, e.g., the $\left\{\tau_i\right\}$ epochs.
The Bayes factor makes the comparison, which is the ratio of the probabilities of the measurements in the two hypotheses.
For independent measurements, we can calculate separate Bayes factors for the position and time measurements, and the combination of the two is just their product, 
\iftrue
\begin{equation}
B = \frac{p(D|H)}{p(D|K)} 
  = \frac{p(D_t|H)}{p(D_t|K)} \frac{p(D_x|H)}{p(D_x|K)}
  = B_t B_x 
\end{equation}
\else
\begin{equation}
B = \frac{L(H)}{L(K)} 
  = \frac{L_t(H)}{L_t(K)} \frac{L_x(H)}{L_x(K)}
  = B_t B_x 
\end{equation}
\fi
The $B_x$ term is the same as for the static objects as discussed in detail by \citet{pxid}. 
The numerator and the denominator of $B_t$ are integrals of the known likelihood functions multiplied by the prior,
%
\begin{eqnarray}
p(D_t|H) &=& \int\!d\tau \ p(\tau|H)\,\prod_i^n L_i(\tau|t_i) \\
p(D_t|K) &=& \prod_i^n \int\!d\tau_i \ p(\tau_i|K)\ L_i(\tau_i|t_i) 
\end{eqnarray}
The choice of the prior naturally affects the Bayes factor, hence it requires some more consideration.

Common sense dictates that the prior on the epoch $\tau$ should be flat as the event can occur at any time with the same probability density. The calculation of the Bayes factor, however, requires a proper prior, whose integral is unity.
Considering that all relevant measurements and the non-zero support of their uncertainties can be constrained to a finite interval without any practical limitations, we can introduce a flat prior on an arbitrary but wide enough \mbox{$[T,\,T\!+\!\Delta]$} interval, and define
\begin{equation} 
\label{eq:const}
p(\tau|\,\cdot\,) =  \left\{\begin{array}{c l}
           \Delta^{-1}  & \quad \mbox{if\ $\tau \in 								[T,\,T\!+\!\Delta]$}\\
           0 & \quad \mbox{otherwise}\\ \end{array} \right.
\end{equation}
Using this flat prior, the likelihoods of the hypotheses become
\begin{eqnarray}
p(D_t|H) &=& \Delta^{-1} \int\!\!d\tau\,\prod_i^n L_i(\tau|t_i) \\
p(D_t|K) &=& \Delta^{-n} \prod_i^n \int\!\!d\tau_i\ L_i(\tau_i|t_i) 
\end{eqnarray}
where we omitted the integration limits as they are irrelevant for any large enough interval, and arrive at the dependence of
\begin{equation}
B_t \propto \Delta^{n-1}
\end{equation}
This means that the wider we make the interval, the larger the Bayes factor becomes, which seems to artificially go against our expectation for an objective quality measure of the associations.

\section{Streams of Events} 
\label{sec:prob}
\noindent
Cosmic events should not be looked at in isolation. A given telescope provides an entire {\em{}stream} of events, which is typically published online in an automated fashion. Our goal is to merge several independent streams of many surveys. The resulting combined stream will carry more information on the individual events, which would help with their scientific analysis. Along with our stream of associations we want to include the probabilities.

The rate of events varies from survey to survey. For a given time interval the number of occurrences can be quite different. Indeed this is a key piece of information that enters the calculations.
By definition the posterior probability of an association is computed from the Bayes factor and the prior of the hypothesis. 
If we assume that the prior probability is uniform, i.e., it takes the same value independent of the position in the sky, etc., we can write it as the ratio
\begin{equation}
P_0 = \frac{N_{\star}}{\prod N_i}
\end{equation}
just like in the static case. 
Here $N_i$ is the number of events in the $i$th survey in a given time interval $\Delta$, and $N_{\star}$ is the number of events that appear in all surveys, i.e., the intersection of their selection functions. 
In our case it is only natural to use $\nu$ frequencies instead, and substitute \mbox{$N=\nu\Delta$} into the prior,
\begin{equation}
P_0 = \frac{\nu_{\star}}{\prod \nu_i}\,\Delta^{1-n} 
\end{equation}
The probability shrinks with increasing time intervals, i.e., it is less probable to randomly select the same event from larger sets. 
In practice, the number of events is always large, otherwise their frequency could not be measured. More formally, $\Delta$ can be arbitrarily wide to yield large $N_i$ counts and a small prior.
For vanishing priors, the posterior depends only on the product of the $P_0$ prior and the Bayes factor
\begin{equation}
P \simeq \frac{P_0 B}{1+P_0 B}
\end{equation}
hence the dependence on $\Delta$ actually cancels out in the product of
\begin{equation}
P_0 B_t = \left( \frac{\nu_{\star}}{\prod \nu_i} \right)\!\!
	    \left[ \frac{\int\!\!d\tau \prod L_i(\tau|t_i)}{\prod \int\!\!d\tau_i\  L_i(\tau_i|t_i)} \right]
	= \pi_0 \beta_t
\end{equation}
The second term of the product, $\beta_t$, can be directly calculated from the data. The value of $\nu_{\star}$ in $\pi_0$ can also be estimated over time from the ensemble statistics of the streams.
We recall the self-consistency argument of \citet{pxid}: the sum of the priors over the all possible $\prod\!N_i$ combinations, which is $N_{\star}$ in total by definition, is also equal to the sum of the posteriors. This equality is an equation for $N_{\star}$, or $\nu_{\star}$ in this case, that in turn provides well-determined prior and posterior probabilities.
In summary, the posterior is well-defined and calculated as
\begin{equation}\label{eq:probx}
P \simeq \frac{\pi_0 \beta_t B_x}{1+\pi_0 \beta_t B_x}
\end{equation}

To select candidate associations early on, when not enough statistics are available to reliably solve for $\nu_{\star}$, one can use the upper bound of \mbox{$\nu_{\star}^{\max}=\min\{\nu_i\}$} to overestimate the posteriors. Similarly if $\nu_i$ are uncertain, one can use conservative limits instead.
By safely overestimating the prior, we end up with somewhat larger posteriors than the actuals, thus a fix probability cut yields slightly more candidates that we can later prune.
This will not overshoot by too much because large posteriors are not particularly sensitive to small variations in the prior \citep{heinis}.

\section{Constraints from Lightcurves}
\label{sec:adv}
\noindent 
One cannot directly measure the epoch of an event, only the fluxes as a function of time. If we can model the change, these lightcurves can provide the time constraints. 
For example, a Type {\sc{}i}a supernova is described by its absolute magnitude $M$, redshift $z$ and epoch $\tau$. For any given set of such parameters we can derive the fluxes at the times of pre-scheduled observations, and compare them to reality.
The likelihood function is given by the deviation of the photometric measurements $f$ and the simulated fluxes $\bar{f}(\tau,M,z)$ as
\begin{equation}\label{eq:like}
L(\tau,M,z|f)=N(f|\bar{f}(\tau,M,z),\Sigma)
\end{equation}
where $\Sigma$ represents the photometric errors in the normal distribution $N$.
By integrating over $M$ and $z$ with their proper priors, we can derive the $\tau$ dependence needed for $\beta_t$.

We use a Gaussian luminosity function \citep{dahlen04} with $M_{\star}$=--19 and $\sigma_M$=0.5 for the prior, and simulate a typical $M_{\star}$ \snIa{} using the LSST cadence \citep{ivezic08}.
In this cadence, pairs of observations separated by an hour are repeated every 3 days. 
Throughout this model we use precision LSST lightcurve templates from Peter Nugent (via Andy Connolly, private communication) but consistently neglect the \mbox{K-correction}. 
The error on the $g$-band photometry is determined from the 5$\sigma$ limiting magnitude of 25. The simulated event is set to happen at 0 day.
The luminosity function determines the kind of \snIa{} events that can occur, and the decline at high redshifts is set by the limiting magnitude of the survey. Together these set the integration domain for our calculations.
The peak brightness of the simulated supernova at redshift $z$=0.8 is 24.5 magnitudes in the $g$ band. This distance is approximately at the peak of the redshift distribution.

\begin{figure}
\epsscale{1.1}
\plotone{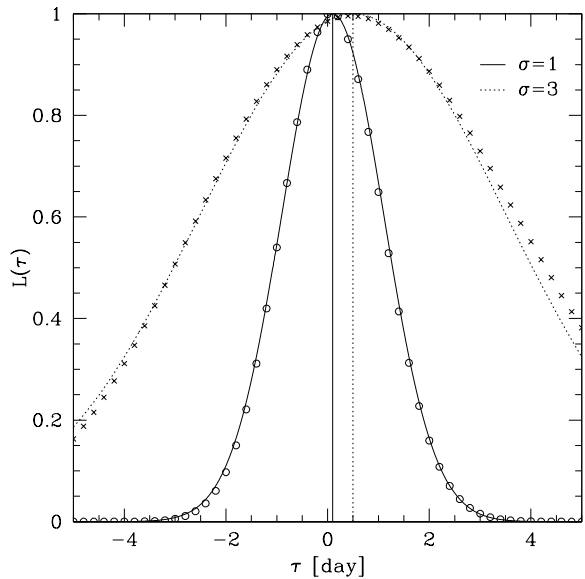}
\caption{The likelihood as a function the epoch is calculated from the lightcurve of a simulated \snIa, whose observations are in the LSST cadence. The {\em{}solid line} is a Gaussian that follows the computed {\em{}open circles} well. When the survey detects a $5\sigma$ transient, it issues an alert. The {\em{}crosses} and the {\em{}dotted line} show the same curve for a truncated lightcurve that is only available up until 2 detections are over the threshold. The shorter baseline yields a weaker constraint.}
\label{fig:like}
\end{figure}

Figure\,\ref{fig:like} shows the likelihood as a function of the epoch for a noiseless set of observations. The results are not sensitive to the end date as long as the integration limit is in the far tail of the decay.
The measurements ({\em{}open circles}) follow a Gaussian surprisingly well. The {\em{}solid line} helps to guide the eye: the curve has $\sigma$=1 day, and a mean of 0.1 days.
Random realizations of lightcurves with realistic noise yield similar profiles with maxima, whose distribution is consistent with the error.
For alerting the transient community, the convention is to have 2 independent observations that pass the detection threshold. For LSST-like cadences, these most likely will happen on the same night 1 hour apart.
As part of the notification message, a lightcurve can be published that consists of the measurements up to the last two 5$\sigma$ observations.
The {\em{}crosses} show the results for such truncated lightcurves, which also closely follow a Gaussian with $\sigma$=3 days and a 0.5 day mean ({\em{}dotted} line). The shorter time baseline and the marginal photometry provide significantly less leverage on the time constraint. We also see more deviation, and a slight asymmetry in the points.
The offset in the peak at these faint magnitudes is a property of the lightcurve's shape  and the cadence. When the photometric errors are tighter, the shift (and the width) is (are) significantly smaller. In comparison, for a $z$=0.5 supernova (with the same $M_{\star}$ luminosity) the brighter lightcurve yields a negligible offset and a $\sigma$$\sim$0.25 days.
More frequent sampling also decreases the offset but it is not practical in real life. A better way is to calibrate the effect and correct for the shifts, if needed.

\begin{figure*}
\epsscale{0.99}
\centering
\plottwo{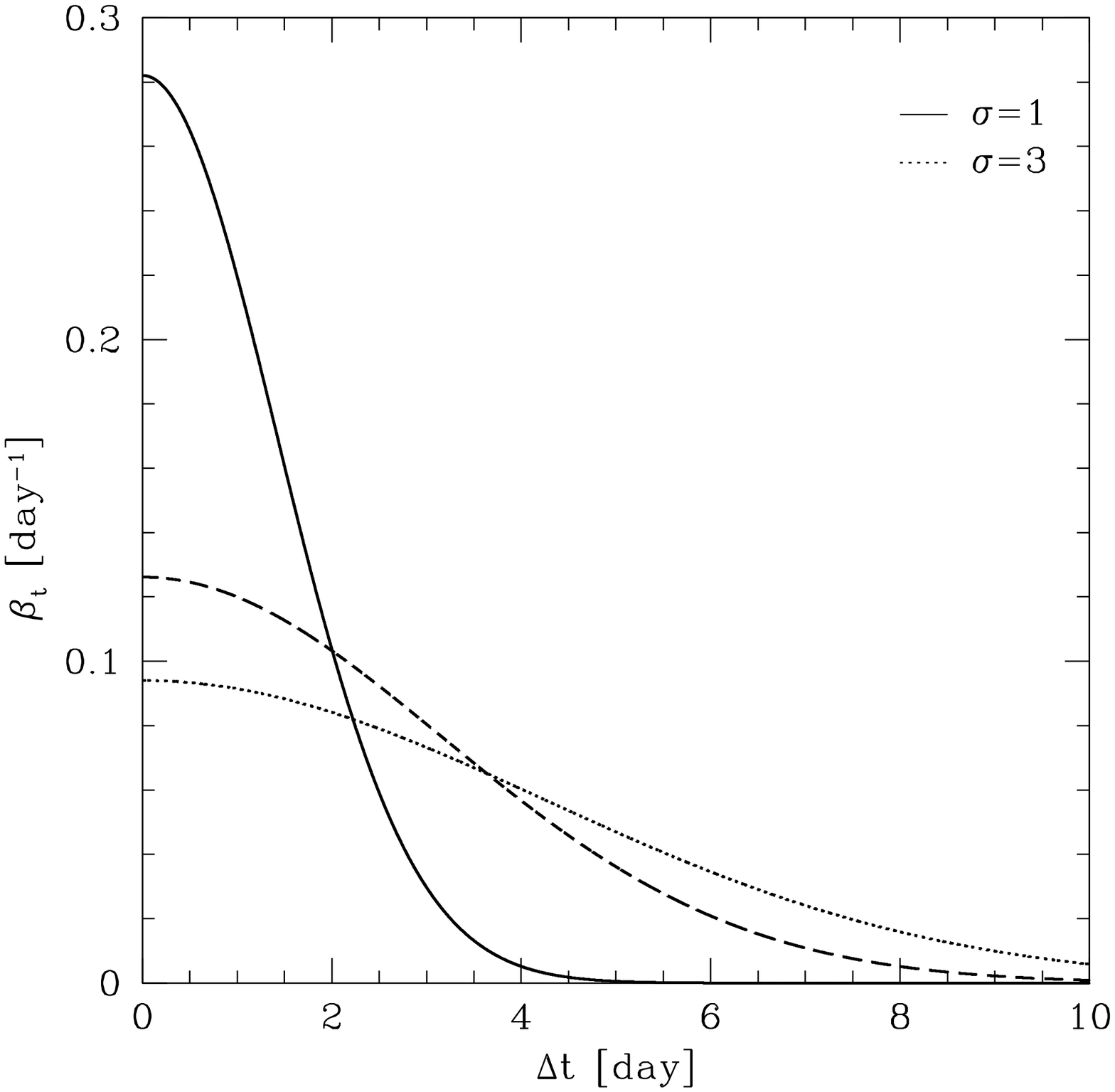}{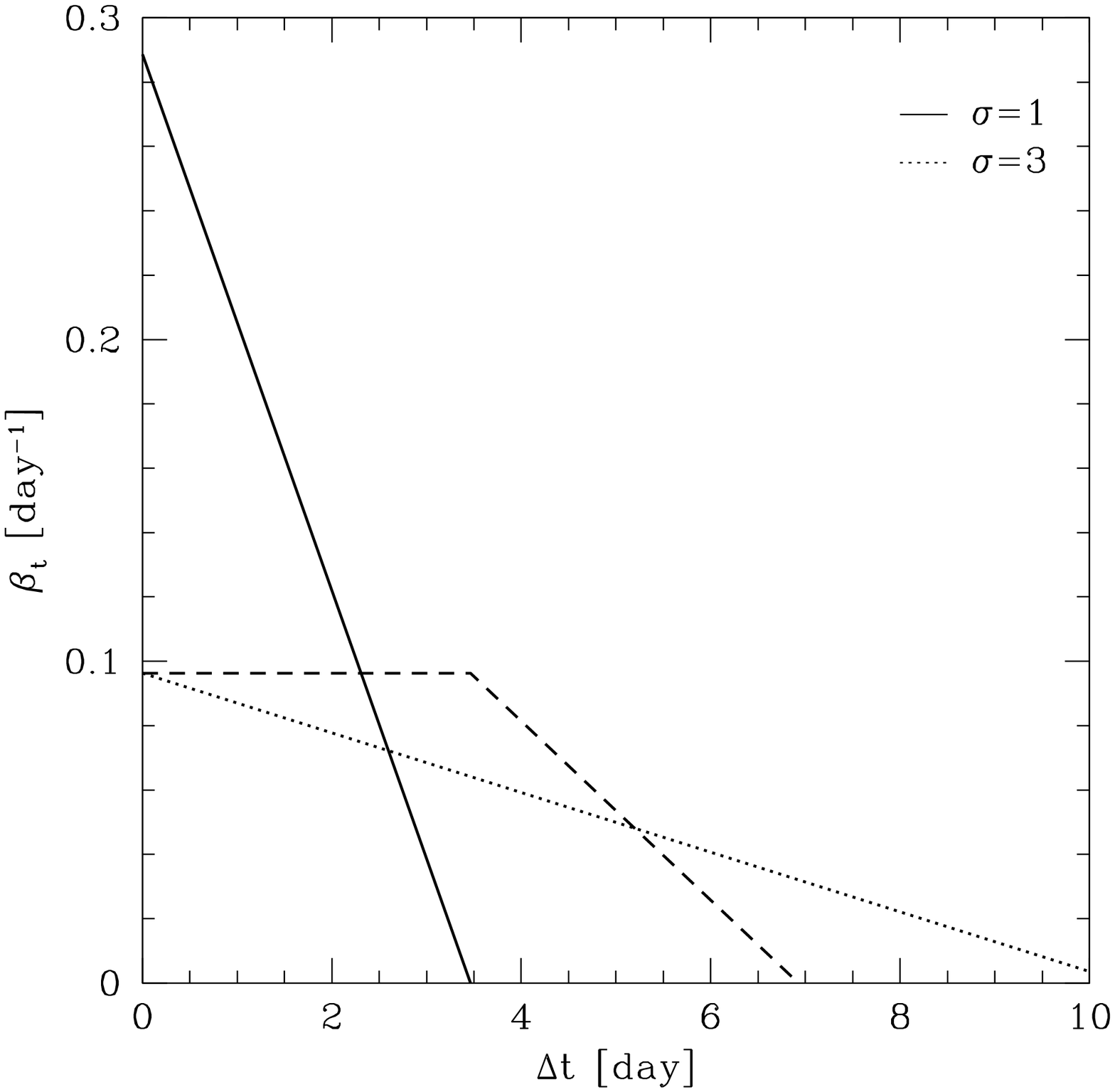}
\caption{When crossmatching streams of events, the $\beta_t$ parameter is the time-domain analog of the Bayes factor. Its dependence on the time difference is plotted for the 2-way Gaussian case in the {\em{}left} panel with uncertainties seen in Figure\,\ref{fig:like}. The line styles are identical to those in Figure\,\ref{fig:like}. The {\em{}dashed line} is for matching one against the other. For comparison, the {\em{}right} panel shows the same using tophat likelihood approximations with the same standard deviations. The finite support of these functions means that $\beta_t$=0/day at large separations in time, which in turn yields 0 posterior probability. The loss of associations, in this case, is a cause for concern, if the width of the time-window is not well-understood.}
\label{fig:beta}
\end{figure*}

%
%
Using Gaussian likelihoods the integrals in $\beta_t$ can be calculated analytically. For a 2-way association it becomes
\begin{equation}
\beta_t = \frac{1}{ \sqrt{2\pi\left(\sigma_1^2+\sigma_2^2\right)} }
\exp\left\{-\frac{(t_1-t_2)^2}{2\left(\sigma_1^2+\sigma_2^2\right)}\right\}
\end{equation}
This is not a Bayes factor and is not dimensionless. 
Substituting $\sigma$=1 day and $\Delta$t=0 yields $\beta_t$=1/$\sqrt{4\pi}$ day$^{-1}$, which is 0.282/day approximately. The numerical integral of the photometric likelihood gives the similar 0.287/day value, which is in good agreement with the analytical estimate. 
The {\em{}left} panel of Figure\,\ref{fig:beta} illustrates $\beta_t$ as a function of $\Delta$t for Gaussians with the aforementioned $\sigma$ values. The {\em{}solid line} shows results from matching with the complete lightcurves and the {\em{}dotted line} is for alerts with truncated photometry. The {\em{}dashed line} illustrates the case of matching a full lightcurve versus an alert.
The {\em{}right} panel shows the same results for a case, when the likelihoods are approximated by tophat functions with the same standard deviations. The finite support of these functions results in $\beta_t$=0/day at large separations in time, which is a cause for concern if the width of the window is not well-understood.
For a $z$=0.5 redshift $M_{\star}$ \snIa{} with $\sigma$=0.25 days the value is 4 times larger, $\beta_t$=1.128/day, when seen at the same time. 

These values of $\beta_t$ are not large, especially considering that $\pi_0$ is typically fairly small, because the time constraint are weak.
The frequency of all events from a telescope can be tens of thousands, say up to $\nu\!\sim$100,000/day, when including all transients. Since the true \snIa{} rate is expected to be around $\nu_{\star}\!\sim$1,000/day in LSST, our estimate for the lower bound comes to $\pi_0\!\sim\!10^{-7}$ days. 
Nevertheless even for the fainter sources, we have higher $P_t\!\sim\!10^{-8}$ values than in the blind matching of static sources, where $P_0\!\sim\!10^{-10}$ \citep{heinis}.
We can increase our prior by filtering the input streams, however, the filtering might be better done on the associations to find the low significance events in the merged stream. In practice, filtering will be applied both before and after for optimal performance.

For completeness we mention that the general result for the \mbox{$n$-way} Gaussian case is also calculated analytically. The resulting formula is similarly simple and reads
\begin{equation}
\beta_t = \left(2\pi\right)^{\frac{1-n}{2}}\!\sqrt{\frac{\prod{}w_i}{\sum{}w_i}}
	\exp\left\{\frac{\left(\sum{}w_i t_i\right)^2}{2\sum{}w_i}\!-\!\frac{\sum{}w_i t_i^2}{2}\right\}
\end{equation}
where $w_i$=$1/\sigma_i^2$ are the precision parameters of the $t_i$ epochs. 

%
%
Photometric measurements can also be directly used to calculate $\beta_t$ using the multivariate likelihood function in eq.(\ref{eq:like}). In addition to a common epoch, identical events have the same physical properties as well, e.g., the absolute magnitude and the redshift of the supernova. Our hypotheses $H$ and $K$ remain the same as before, but we parameterize them with not only $\tau$, but also with $M$ and $z$. To calculation of $\beta_t$ involves computing the integrals over all three parameters.
Of course the numerical integral is substantially more expensive computationally than evaluating an analytic formula 
but this is the proper use of the model that fully exploits its constraining power.
Following the same prescription as before, we generated two lightcurves using the LSST cadences. The only difference was in their phase: the cycle of one simulated survey was shifted by a day from the other.
The result for an $M_{\star}$ supernova at $z$=0.8 redshifts is almost 20 times larger than just based on time coincidence, $\beta_t$=5.3/day.
If we are willing to live with the extra computation and the explicit dependence on a model, this way we can boost the probability, in this case, to almost 1\%. Naturally, one would still need spatial constraints for any meaningful match, but could potentially leverage lower accuracy positional measurements.

The explicit modeling of the lightcurve provides an opportunity to simultaneously perform classifications and outlier detection. We can benefit from calculating several Bayes factors or $\beta_t$ values for comparison. 
Associations that appear to be reliable matches based on their directions and epochs when using a model independent approximate time window, might be rejected based in a detailed analyses of their SEDs. Accidental noise or new discovery? Follow up observations might help to decide. 
Beyond the above outlier detection, we can also run several SED models to find out in which class the candidate event might fit best. 
Naturally there is also nothing to stop us from creating several merged streams using different strategies, which target specific scientific experiments and communities.

\section{Conclusions}
\label{sec:sum}
\noindent
We introduced a new statistical method to associate cosmic events based on measured constraints in both time and space. Folding in the time-domain information is important for a number of reasons: 
(1)
If we can automatically cross-correlate streams of events from multiple surveys, we can find fainter ones that are not obvious in any one of the observations but together become significant. This enables us to push these studies to uncharted territories. 
(2)
Constraints on the epoch can boost small positional evidences in case of low accuracies, e.g., gamma-ray bursts.
(3)
A more fundamental, yet, extremely important practical matter is the assignment of probabilities. 
While we can clearly calculate Bayes factors based on positional information only, we cannot define the probabilities without the time-domain data.
Conceptually the problem is with the prior. It cannot be determined without counting the occurrences of events in a fixed time interval, whose duration is formally arbitrary. We found that the solution is to fold in the Bayes factor in time, which naturally cancels the artifact.

The lightcurve measurements constrain the epoch of a given event. For Type {\sc{}i}a supernovae, we found the likelihood to be very close to a Gaussian that makes the problem analytically tractable. The numerical simulations agree with the fast approximations. 
The same method also works for streams of events seen in different wavelengths as long as regime is covered by the SED models. 
To unlock the full potential of the lightcurves one can further exploit the SED modeling.
The calculation is extended to require matching events to have the same physical properties. For \snIa{}, the numerical integrals show that this strategy has the potential to boost the time-domain evidence by a factor of 20, which means that one can get away with less accurate spatial measurements.

Although we illustrated the methodology on supernovae, it is clearly applicable to other types of events. 
Before gamma-ray bursts were established to be extragalactic, \citet{bf_grb} applied Bayesian hypothesis testing to look for repeating gamma-ray bursts. Since then we have seen the host galaxy of many GRBs. Using the time-domain cross-identification we can look for other most subtle events in their hosts from related physical processes.
Another example is the detection of gravitational waves. The LIGO Collaboration discusses astrophysically triggered searches \citep{ligo}. Their idea is to lower the detection threshold for a short period of time, when other type of cosmic events occur. This is in fact a very similar problem: merging streams of events from LIGO and other projects with our probabilistic approach could boost their significance and help facilitate new discoveries.

The general cross-identification problem in astronomy is rather convoluted and ultimately tightly interwoven with the scientific analyses. One obvious example is classification, which in turn would affect the matching prior. 
The circular nature of the problem is not an issue but one needs to find a consistent solution.
The explicit assumptions in the probabilistic approach enables us to incorporate new, different kind of data on top of the epochs and the positions, when available.
This is a crucial feature, which will prove vital in studying the time-domain and understanding the eventful Universe.

\acknowledgements 
The author would like to thank Andy Connolly, Alex Szalay, Istv\'an Csabai, Matthew Graham, Suvi Gezari, Josh Bloom, Joey Richards, Roy Williams and Tom Loredo for several stimulating discussions on various aspects of the topic.
This study was supported by the Gordon and Betty Moore foundation via GBMF 554.


\begin{thebibliography}{}

\bibitem[Abbot et~al.(2008)]{ligo} Abbott, B., et~al.\ 2008, Class. Quantum Grav., 25 114051

\bibitem[Budav\'ari \& Szalay(2008)]{pxid} Budav\'ari, T., \& Szalay, A.~S., 2008, \apj,  679, 301

\bibitem[Dahlen et~al.(2004)]{dahlen04} Dahlen, T., et al.\ 2004, \apj, 613, 189 


\bibitem[Drake et~al.(2009)]{crts} Drake, A.~J., et~al., 2009, \apj, 696, 870

\bibitem[Heinis et~al.(2009)]{heinis} Heinis, S., Budav{\'a}ri, T., \& Szalay, A.~S.\ 2009, \apj, 705, 739

\bibitem[Ivezic et~al.(2008)]{ivezic08} Ivezic, Z., et al.\ 2008, Serbian Astronomical Journal, 176, 1 

\bibitem[Keller et~al.(2007)]{skymapper} Keller, S.~C., et~al.\ 2007, Publications of the Astronomical Society of Australia, 24, 1


\bibitem[Luo et~al.(1996)]{bf_grb} Luo, S., Loredo, T., \& Wasserman, I.\ 1996, American Institute of Physics Conference Series, 384, 477 

\bibitem[Rau et~al.(2009)]{ptf} Rau, A., et al.\ 2009, \pasp, 121, 1334 






\end{thebibliography}
\end{document}